\documentclass[]{emulateapj}

\newcommand{\be}{\begin{equation}}

\newcommand{\ee}{\end{equation}}

\newcommand{\lavec}[1]{{\mathbf{#1}}}         

\newcommand{\Rn}{{\sf {R_v}}}

\def\al{Alfv\'en\ }

    \def\u{{\lavec{u}}}     

\slugcomment{Accepted for publication in Astrophysical Journal}

\shorttitle{Solar Coronal Heating}
\shortauthors{Ng, Lin, and Bhattacharjee}

\begin{document}


\title{High-Lundquist Number Scaling in Three-Dimensional Simulations of
  Parker's Model of Coronal Heating}

\author{C. S. Ng}
\affil{Geophysical Institute, University of Alaska Fairbanks, PO Box 757320, Fairbanks, AK 99775, USA}
\email{chung-sang.ng@gi.alaska.edu}

\author{L. Lin and A. Bhattacharjee}
\affil{Space Science Center, University of New Hampshire, 39 College Road, Durham, NH 03824, USA}




 
\begin{abstract}
Parker's model is one of the most discussed mechanisms for coronal heating and has generated much debate. 
We have recently obtained new scaling results in a two-dimensional (2D) version of this problem 
  suggesting that the heating rate becomes independent of resistivity in a statistical steady state 
  [Ng and Bhattacharjee, Astrophys. J., {\bf 675}, 899 (2008)]. 
Our numerical work has now been extended to 3D by means of large-scale numerical simulations. 
Random photospheric footpoint motion is applied for a time much longer than the correlation time 
  of the motion to obtain converged average coronal heating rates. 
Simulations are done for different values of the Lundquist number to determine scaling. 
In the high-Lundquist number limit, 
  the coronal heating rate obtained so far is consistent with a trend that is independent of the Lundquist number, 
  as predicted by previous analysis as well as 2D simulations. 
In the same limit the average magnetic energy built up by the random footpoint motion tends to 
  have a much weaker dependence on the Lundquist number than that in the 2D simulations,
  due to the formation of strong current layers and subsequent disruption when
  the equilibrium becomes unstable. We will present scaling analysis showing that
  when the dissipation time is comparable or larger than the correlation time of the random
  footpoint motion, the heating rate tends to become independent of Lundquist number, 
  and that the magnetic energy production is also reduced significantly.
\end{abstract}


\keywords{magnetic reconnection --- magnetohydrodynamics (MHD) --- Sun: corona --- Sun: magnetic topology}

\section{Introduction}

The enormous energy content of high-beta photospheric plasma flows has long been suggested
as the source of energy that ultimately heats the million degree solar
corona. Unambiguously identifying the exact mechanisms that transfer this
kinetic energy into the overlying solar atmosphere and the exact nature of how the coronal
magnetic field responds and converts this energy into heat remains one
of the long-standing issues in astrophysics.

In this paper we investigate an idealized model of the corona 
proposed by \cite{Parker1972} which applies to closed magnetic field
structures whose field lines are embedded at both ends in the solar surface.
Neglecting the curvature of the magnetic field, the corona is modeled in Cartesian geometry where
an initially uniform magnetic field along the $\hat{e}_{z}$
direction is line-tied at $z=0$ and $z=L$ in perfectly conducting
end-plates representing the photosphere. 
Parker suggests that slow and continuous random shuffling of the
footpoints at these end-plates, representing the turbulent buffeting of the
coronal field embedded in the convecting photosphere,
can tangle the field into a braided structure of sufficient complexity such
that it cannot settle into a continuous smooth equilibrium, but rather
necessarily evolves to one with tangential discontinuities.
Whether or not true current singularities (as opposed to current layers
with finite thickness) can form in this scenario and whether or not continuous
footpoint mappings necessarily imply a non-smooth topology has been the
subject of intense debate in the decades that have passed since Parker's
seminal proposal. Extended discussion of this matter is beyond the scope of the
present analysis, but it is appropriate to reiterate here (c.f. \citealt{nb1998})
that this question is not merely of academic interest. That the plasma
gradients will tend towards singularities has important bearing on the physics
of magnetic reconnection and turbulence dynamics in the corona. The interested
reader is referred to \cite{nb1998}, \cite{Low2010}, \cite{HBZ2010PhPl},
\cite{JanseLowParker2010} and references therein. 

In a process Parker calls ``topological dissipation'', it is at these
tangential discontinuities where the corona's small but
ultimately finite resistivity induce the formation of current sheets where
magnetic energy is dissipated to heat the coronal plasma, and where magnetic
reconnection proceeds to reduce the topological complexity of the coronal
magnetic field. This essential concept was further developed in a series of 
studies (\citealt{Parker1979, Parker1983a, Parker1983b, Parker1988, Parker1994})  and has become known colloquially
as the ``nanoflare model'' of coronal heating. The appellation derives
from the isolation of $10^{23}$ erg flares as the constitutive energy release
events which occur in ``storms'' of sufficient ferocity to heat the corona and
adequately account for observed conductive and radiative losses.  While the
concept of topological dissipation can be seen as the prototypical ``DC''
mechanism for coronal heating (c.f. \citealt{Klimchuk2006,Aschwanden2005}), the solar atmosphere surely admits more complex
magnetic topology than is treated by the Parker model. In fact, many
investigators have pursued reconnection-based heating mechanisms using
geometries that include separators, separatrices and magnetic-nulls (see
\citealt{Priest2005}, and references therein), and more recently by analyzing
the magnetic topology  of active regions observed by {\it Transition Region and Coronal Explorer} \citep{Lee2010ApJ}. 
It remains clear however, that coronal loops are the  basic
building block of the solar corona, and that their examination first as isolated
entities is crucial in laying the foundations for a broader understanding
of the corona and its activity (see \citealt{Reale2010LRSP} for a recent review).

The work we present here is motivated by a recent study 
\citep[hereafter NB08]{nb2008} which developed a simplified version of the Parker scenario.
NB08 referred to this version as a constrained tectonics model \citep{PHT2002} because 
the random braiding at the line-tied ends was restricted to
depend on only one coordinate transverse to the initial magnetic field. 
This strong assumption has the consequence that it enables us to describe 
the complete dynamics of the system by a simple set of differential equations
which are easily amenable to analytical and numerical solutions for
prescribed footpoint motion. 
The geometric constraints imposed by our assumption preclude the occurrence
of nonlinear effects such as reconnection and secondary instabilities, 
but enables us to follow for long times the dissipation of energy due to the
effects of resistivity and viscosity. Using  this model, it was shown both
numerically and by scaling analysis that as long as the correlation time of
turbulent photospheric flow ($\tau_c$) is much smaller than the characteristic
resistive timescales ($\tau_R$), ohmic  dissipation becomes independent of
resistivity ($\eta$). The absence of nonlinear effects in this model
allows the perpendicular magnetic field ($B_{\perp}$) to grow to unphysically
large values and is found to scale as $\eta^{-1/2}$.  Furthermore, NB08
conjectured, by means of a heuristic scaling argument, that even in the
presence of reconnection and secondary instabilities, the heating rate would
remain insensitive to resistivity. It is this conjecture that we examine here using
three-dimensional (3D) reduced MHD (RMHD) simulations.

The Parker model has been studied extensively using 3D MHD
numerical simulations
(\citealt{Mikic1989,ls1994,Einaudi1996,Hendrix1996b,GalsNord1996,Dmitruk1998,Gomez2000,Rappazzo,Rappazzo2010a}
among others.) Here we are interested
in the precise scaling of dissipation with respect to plasma resistivity. Our
study is most similar in design to that of \cite{ls1994} who
used RMHD to simulate Parker's model with Lundquist
numbers spanning an order or magnitude. In this range they found that both
heating rate and perpendicular field production scales as $\eta^{-1/3}$.
These numerical results agreed with analysis based on Sweet--Parker reconnection theory
and measurements of current sheet statistics.

In this paper, we recover the scalings of heating rate and of $B_{\perp}$ of \cite{ls1994} in
the range they examined, and extend their results to a range of
lower $\eta$, where the results support a slower growth of
$B_{\perp}$ which roughly scales as $\eta^{-1/5}$ and a heating rate that
becomes insensitive to $\eta$.  We also demonstrate by simple scaling analysis
that the transition between these scaling behaviors results from the
diminishing effects of random photospheric motion as the energy dissipation
timescale $\tau_E$ becomes much smaller than the correlation time $\tau_c$,
in accordance with NB08.

The paper is organized as follows. In Section~\ref{NumMod} we describe the
RMHD coronal heating model and numerical scheme.
Section~\ref{NumRes} gives the details of the simulation results.
Section~\ref{LongFlx} presents a scaling analysis describing
the transition in the observed scaling behavior. Section~\ref{fwcr} summarizes our
conclusions to date and discusses outstanding physical issues. 


\section{Coronal Heating Model and Numerical Setup}
\label{NumMod}


We assume that the coronal plasma is sufficiently low-beta so that 
  the dynamics can be described by the RMHD equations.
Many numerical studies of Parker's model mentioned above
(with the notable exceptions of \citealt{GalsNord1996,GudNord05,GudNord06}) are
based on the RMHD approximation.
The RMHD equations are a simplified version of MHD applicable to systems
  where the plasma is dominated by a strong guide field such that 
  the timescales of interest are slow compared with the
  characteristic Alfv\'{e}n timescale $\tau_{A}$. 
These restrictions also imply
  incompressibility ($\nabla\cdot \mathbf{v}=0$) and the exclusion of magnetosonic modes 
  (leaving only the shear Alfv\'{e}n modes propagating in $\hat{e}_{z}$).  
The RMHD equations were first  derived for the
  study of tokamak plasmas by \cite{KadPog1974} and \cite{strauss76},
which can be written in dimensionless form as
\begin{equation}
\frac{ \partial \Omega}{\partial t}+[\phi,\Omega]=\frac{\partial J}{\partial
  z} +[A,J]+ \nu \nabla^2_\perp \Omega
\label{eq_momentum}\end{equation}
\begin{equation}
\frac{ \partial A}{\partial t}+[\phi,A]=\frac{\partial \phi}{\partial
  z} + \eta \nabla^2_\perp A
\label{eq_induction}\end{equation}
where $A$ is the flux function so that the magnetic field is expressed as 
$\mathbf{B=\mathbf{\hat{e}_{z}}+B_\perp}=\mathbf{\hat{e}_{z}}+\nabla_\perp A\times\mathbf{\hat{e}_{z}}$;
$\phi$ is the stream function so that the fluid velocity field is expressed as 
$\mathbf{v}=\nabla_\perp \phi\times\mathbf{\hat{e}_{z}}$;
$\Omega= -\nabla^2_{\perp}\phi$ is the  $z$-component of the vorticity;
$J = -\nabla^2_{\perp}A$ is the  $z$-component of the current density;  
and the  bracketed terms are Poisson brackets such that, for example, $[\phi,A]\equiv\phi_yA_x-\phi_xA_y$
  with subscripts here denoting partial derivatives. 
The normalized viscosity $\nu$ is  the inverse of the Reynolds number $\Rn$, 
and resistivity $\eta$ is  the inverse of the Lundquist number $S$.
The normalization adopted in Equations (\ref{eq_momentum}) and (\ref{eq_induction}) is such that 
  the magnetic field is in the unit of  $B_z$ (assumed to be a constant in RMHD);
velocity is in the unit of $v_A = B_z/(4\pi\rho)^{1/2}$ with a constant density $\rho$; 
length is in the unit of the transverse length scale $L_\perp$; 
time $t$ is in the unit of $L_\perp / v_A$;
$\eta$  is in the unit of $4\pi v_A L_\perp /c^2$; 
and $\nu$  is in the unit of $\rho v_A L_\perp$.  

In this paper we investigate an idealized model of the corona 
proposed by \cite{Parker1972}.
In Parker's model, a solar coronal loop
is treated as a straight ideal plasma column, 
bounded by two perfectly conducting end-plates representing the photosphere.  
The footpoints of the magnetic field in the photosphere are frozen (line-tied).  
Initially, there is a uniform magnetic field along the $z$  direction.  
The footpoints on the plates $z = 0$ and $z = L$ are subjected to slow, 
random motion $\phi(z=0,t)$ and $\phi(z=L,t)$ that deform the magnetic field. 
The method we follow for imposing random footpoint motion is described in detail
in \cite{Long1993PhD}. We use a correlation time of $\tau_c = 10$. Only a band of
Fourier modes with small $k$ is driven so that the boundary flow is of
the characteristic length scale of the order of the perpendicular simulation
box size. The amplitude of the driving is small enough such that the
root-mean-square boundary flow ($v_{\rm rms} \sim 0.075$) is small compared
with the \al speed along the large-scale magnetic field. 

The footpoint motion is assumed to take place on a timescale much longer 
  than the characteristic time for \al wave propagation between $z = 0$  and $z = L$, 
so that the plasma can be assumed to be in quasi-static equilibrium nearly everywhere, 
if such equilibrium exists, during this random evolution.  
For a given equilibrium, 
a footpoint mapping can be defined by following field lines from one plate to the other.  
Since the plasma is assumed to obey the ideal MHD equations, 
the magnetic field lines are frozen in the plasma and cannot be broken during the twisting process.  
Therefore, the footpoint mapping must be continuous for smooth footpoint motion.  
Parker \citep{Parker1972} claimed that if a sequence of random footpoint motion 
  renders the mapping sufficiently complicated, 
there will be no smooth equilibrium for the plasma to relax to, 
and tangential discontinuities (or current sheets) of the magnetic field must develop.
Although Parker's claim has stimulated considerable debate, (see
\citealt{JanseLowParker2010}),
we have shown elsewhere that it is valid if the equilibrium becomes unstable 
  because there is only one smooth equilibrium for a given footpoint mapping \citep{nb1998} 
  under RMHD and using periodic boundary conditions in transverse coordinates.
Moreover, thin current layers generally do appear in numerical simulations
  with small but finite resistivity,
after random boundary flows have been applied for a period of time.

Ordinarily, the problem of calculating time-dependent solutions of Equations 
  (\ref{eq_momentum}) and (\ref{eq_induction}) in line-tied magnetic field geometry involves 
  all three spatial coordinates and time. 
As a first step in NB08, 
we had made a strong assumption that in addition to the coordinate $z$ 
  along which the magnetic field is line-tied, 
the dynamics depends on only one transverse coordinate $x$ (as well as  time $t$).  
The nonlinear terms (those that involve Poisson brackets) in RMHD equations 
  (\ref{eq_momentum}) and (\ref{eq_induction}) then become identically zero.
  
We have developed computer simulation codes that integrate Equations 
  (\ref{eq_momentum}) and (\ref{eq_induction}) numerically for arbitrary 
  footpoint displacements in both 2D and 3D. 
We use spectral decomposition in $x$ and $y$ 
  (using a standard two-thirds de-aliased pseudo-spectral method)
  and a leapfrog finite difference method in $z$ on a staggered grid. 
The accuracy of the Fast Fourier Transform (FFT) routine used has been tested 
  extensively as shown in \cite{Ng-etal-2008}.
For the 2D version, we can use an implicit method for time-integration so that we can take 
  larger time steps than is allowed by the Courant-Friedrich-Lewy (CFL) condition 
  for numerical stability of explicit methods, 
unlike in the 3D version,
  in which a predictor-corrector time-stepping is used. 

In NB08, the 2D code was run for cases with different resistivities. It was found
that if $\tau_c \ll \tau_R$, ($\tau_R$ is the characteristic resistive
diffusion timescale), measured average rates of ohmic and viscous
dissipation became insensitive to resistivity. A simple scaling analysis
showed that this behavior could be derived beginning from general
considerations if one includes the effects of the random walk nature of photospheric footpoint
motions. However, because the simulations lacked instabilities or magnetic
reconnection, the growth of magnetic energy was unbounded,
with average transverse magnetic field $\bar{B}_y$ scaling as
$\eta^{-1/2}$. NB08 went on to demonstrate by further analytical argument that even if 
the growth of the magnetic field was thwarted, as would be the case in 3D
simulations, dissipation would remain independent of resistivity, regardless
of both the specific saturation level of $B_{\perp}$ and of the mechanism 
causing the saturation. It is this conjecture we seek to examine.

The prescribed ordering $\tau_{c} \ll \tau_{E}$ ($\tau_E$ is the
characteristic timescale over which energy is built up before being
impulsively dissipated) is not met for the 3D simulations of
\cite{ls1994}, who found  both dissipation and  $\bar{B}_y$  to scale as
$\eta^{-1/3}$. For our numerical experiment, we extend the range they examined by an order
of magnitude in either direction, establishing an adequate domain to assess the
stated hypothesis. We have also adopted their footpoint braiding algorithm (as
described in \citealt{Long1993PhD}), allowing a direct comparison of 
scaling results in the range they examined. Before
proceeding with the details of our simulations, it is worth noting that NB08
demonstrated analytically as well as numerically
that in the presence of steady footpoint motion, i.e., when $\phi(z=0)$ and
$\phi(z=L)$ are time-independent, the heating rate is inversely proportional to
$\eta$, which is a strong and physically unrealistic dependence. For this
reason, we do not pursue steady boundary flow, used in some other studies of coronal heating
(e.g. \citealt{Rappazzo}).

\section{Numerical Results: Statistical Steady State}
\label{NumRes}

We have performed a series of simulations using our 3D RMHD code
described in Section~\ref{NumMod} 
using a range of $\eta$ spanning two orders of magnitude to study scaling laws.
Extending the range in $\eta$ by an order of magnitude beyond what was studied
by \cite{ls1994} poses a significant
challenge. 
As the dissipation coefficients ($\eta$ and $\nu$) get smaller,
higher resolution has to be used to resolve smaller scales.
To run the 3D code in high resolution, 
   simulations are performed on parallel computers using MPI 
   (Message Passing Interface).
In order to run some cases for even longer time, 
  we have also modified the code and run it on machines with GPUs
  (Graphics Processing Units) using Nvidia's Compute Unified Device
  Architecture (CUDA). 

The range of $\eta$ has been extended to lower values (with $\tau_c  = 10 \ll \tau_r$)
  for about an order of magnitude 
  as compared with the study in \cite{ls1994} which stopped at $\eta \sim 10^{-3}$.
This extension of course requires significant increase in resolution, 
with our highest resolution case at $512^2 \times 64$ so far
(runs with even higher resolution, such as R0 in Table~\ref{tbl-1},
have not been run for long enough time for good statistics), 
as compared with $48^2 \times 10$ in \cite{ls1994}.
The main difficulty in performing these simulations is the requirement to run up to hundreds 
  or even thousands of  \al times in order to obtain good statistics of the average quantities 
  under  the driving of random boundary flow. 
The basic parameters and results in
  this scaling analysis are summarized in Table~\ref{tbl-1}.  

\begin{figure}
\epsscale{1.0}
\plotone{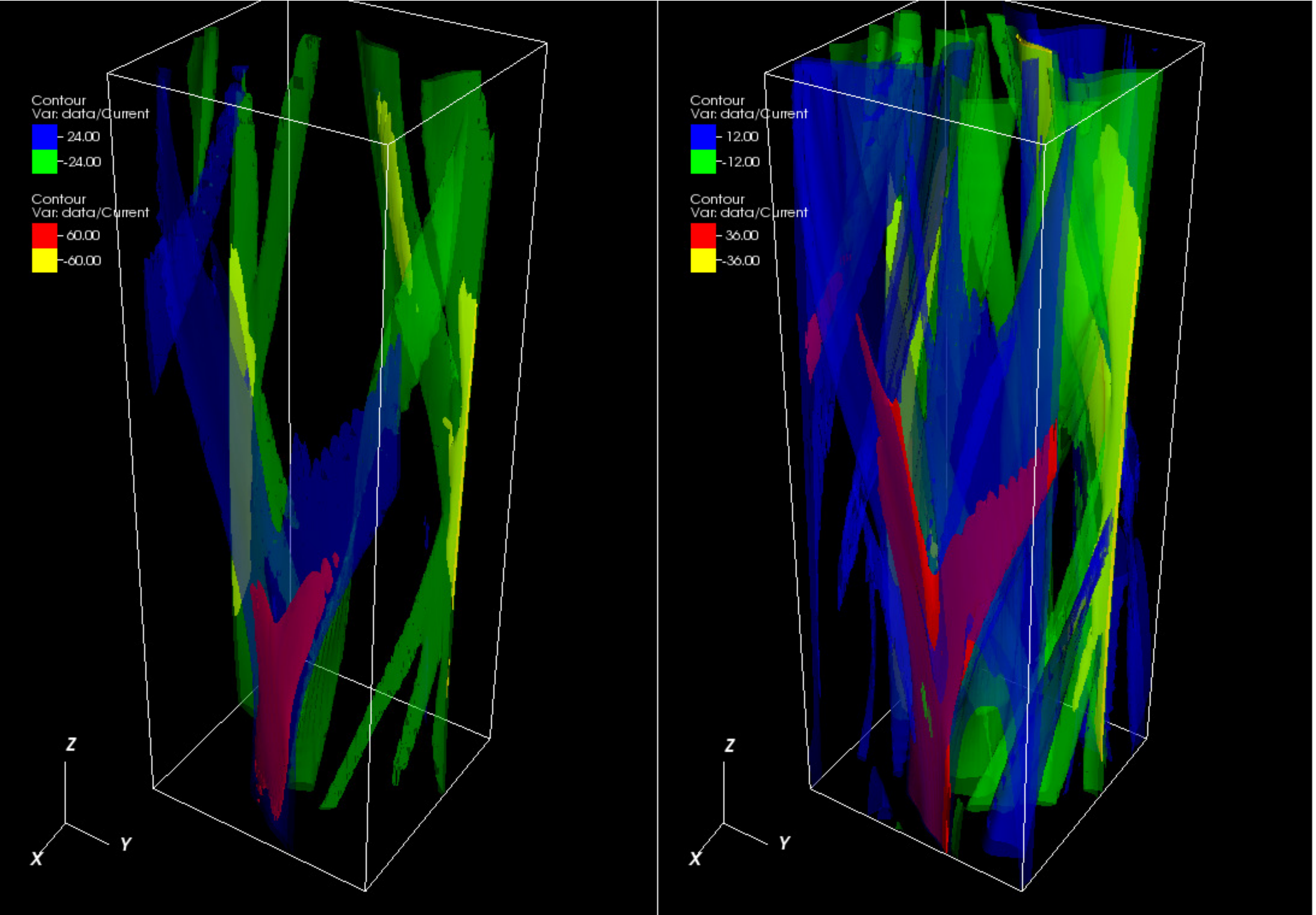}
\caption{3D iso-surfaces of $J$ at a time taken from the R7 run. Both figures
  are from the same time sample for the R7 run. The left panel shows
  iso-surfaces at $J=-60,-24,24,60$ while the right panel shows iso-surfaces
  at $J=-36,-12,12,36$. Blue and green iso-surfaces are made semi-transparent
  for greater visibility. }
\label{cs}
\end{figure}

\begin{figure}[t]
\epsscale{1.0}
\plotone{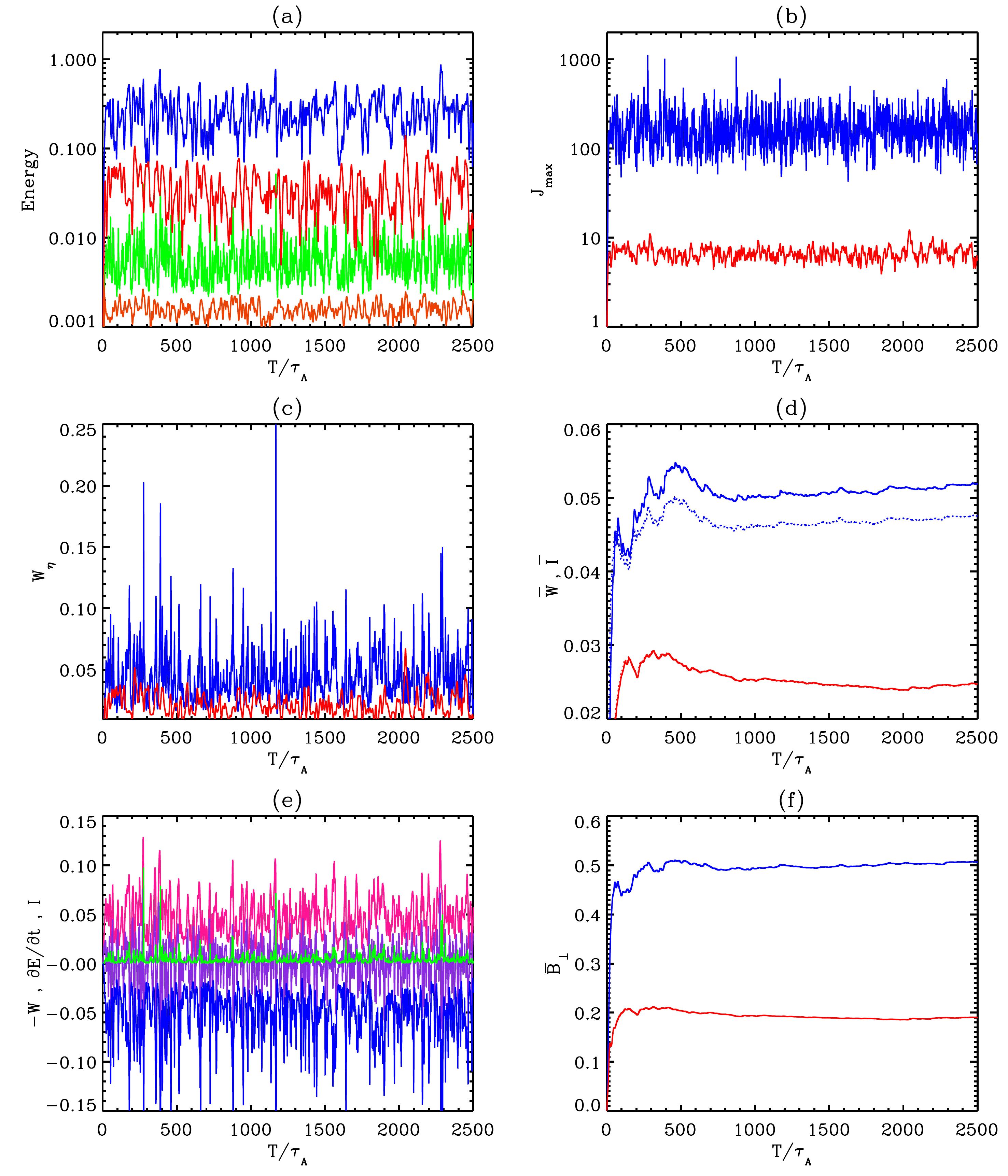}
\caption{Plots (a), (b), (c), (d), and (f) show time series of various quantities for
  runs R5 (Blue) and R12 (Red). In (a), green and orange correspond to
  $E_{K}$ for R5 and R12 respectively while blue and red show
  $E_{M}$. For plot (d), solid lines show $\bar{W}$ and dotted lines show
  $\bar{I}$. For run R5 plot (e) shows $-W=-(W_\eta + W_\nu)$ (Blue), $I$
  (Pink), $d(E_M + E_K)/dt$ (Purple), and the difference between the
  right and left hand sides of Equation~\ref{eng_bal} (Green). Parameters used for
  these runs can be found in Table~\ref{tbl-1}. \label{ts}}
\end{figure}

It is crucial to the scaling study that we obtain good statistics averaging
  over time evolution in statistical steady state. 
As with previous long time integration studies of the Parker model, 
the runs are started with a uniform magnetic field along $\hat{e}_{z}$. 
After initial transients,
the system will evolve to a statistical steady state. 
As mentioned above, thin current layers 
are formed and dissipated repeatedly during this statistical steady state.
Figure~\ref{cs} shows 3D iso-surfaces of $J$ at a time taken from the R5 run,
  when there is a larger number of current sheets.
This process is repeated indefinitely as the random boundary flows 
  keep twisting the magnetic fields.
Energy of the system is dissipated impulsively. 
As Poynting flux
  injection progressively braids the fields, energy is built up until an
  instability  drives current sheet formation and reconnection, after which
  energy is released in a short time.
This is a major characteristic of the statistical steady state.
For the runs R5 (Blue) and R12 (Red), specified in Table~\ref{tbl-1},
Figure~\ref{ts} shows the intermittent nature of various quantities in time
(in the unit of the \al time $\tau_A$, the time it takes for \al waves to travel
the distance of $L = 1$ between the two boundary plates along $z$).

Figure~\ref{ts} (a) shows the total magnetic energy 
  $E_M = \int {\bf B}^2_\perp d^3x$, 
as well as total kinetic energy 
  $E_K = \int {\bf v}^2_\perp d^3x$, 
where the integration is over the 3D simulation box. 
Note that the magnetic energy does not include the contribution from the $B_z$
  component, which is constant in the RMHD model.
Since the applied photospheric flow is chosen to be small (less than one tenth) 
  compared with the \al speed (with $\tau_c  = 10 \tau_A$),
  the magnetic field configuration maintains quasi-equilibrium for most of the time, 
  except when strong current sheets form from time to time, thus inducing instabilities and
  strong dissipation.
Therefore, $E_M$ is usually much larger than $E_K$.

Figure~\ref{ts} (b) shows the maximum current density $J_{\rm max}$
  over the whole 3D volume. 
$J_{\rm max}$ increases over an order of magnitude on average in R5 compared with R12,
  and also fluctuates in time over a much larger amplitude.
Observing Figures~\ref{ts}(a) and \ref{ts}(b), note that the ratio of the increase in $J_{\rm max}$ is much larger
than the ratios of the increases of both $E_M$ and $E_K$  as $\eta$ decreases.

Figure~\ref{ts} (c) shows the ohmic dissipation
  $W_\eta = \eta \int J^2 d^3x$.
Similarly, there is also energy dissipated by viscosity,
 at a rate given by $W_\nu = \nu \int \Omega^2 d^3x$ (not shown).
Due to the same reason that $E_K$ is much smaller than $E_M$,
  the viscous dissipation is  much smaller than the ohmic dissipation,
  if we choose $\nu = \eta$ (Prandtl number equal to unity), which holds for
  most of our simulations.
In this case, the total energy dissipation rate (heating rate) is dominated 
  by ohmic dissipation.
When we use values of $\nu$ greater than $\eta$, the viscous dissipation can become a
more significant fraction of the ohmic dissipation. From the plot of $W_\eta$, we note
that even though it shows large fluctuations in time, it fluctuates around a
higher level for R5 than for R12 due to the smaller value of resistivity in
the former.

 
To give a better measure of the level of energy dissipation, 
we can calculate the time averaged energy dissipation rates, e.g., 
$\bar{W_\eta} =  [ \int^t_0 W_\eta dt' ]/t$, and similarly for $\bar{W_\nu}$.
The total energy dissipation rate is then $\bar{W} = \bar{W_\eta} + \bar{W_\nu}$,
which is plotted in Figure~\ref{ts}(d).
Our physical assumption here is that such averaged quantities will tend 
  to saturated levels as $t$ tends to infinity. 
In practice, since we can only simulate for finite time, 
  such saturated levels are found at a time $t \gg \tau_c \gg \tau_A$
  when these time-averaged values do not fluctuate too much.
We do see from this plot that $\bar{W}$ tends to saturate after a time much 
larger than $\tau_A$.
  
Also plotted in Figure~\ref{ts}(d) is the time average Poynting flux $\bar{I}$, 
  where $I = B_z \int {\bf v} \cdot {\bf B} d^2x$,
integrated over the top and bottom boundary surfaces with ${\bf v} = \u_p$,
  the random photospheric flow.
Note that $I$ is not positive definite due to the fact that it involves the dot product
  between the velocity and magnetic field vectors and thus can be either positive
  or negative.
However, the time averaged $\bar{I}$ is almost always positive due to two factors.
First, due to ohmic and viscous dissipation of energy into heat, 
  if the total energy of the system is at a statistically steady level,
  there must be energy input from the boundary to provide this dissipation loss.
Secondly, even when there is not much energy dissipation during a certain period,
magnetic energy $E_M$ generally increases, since the magnetic footpoints at
  the two boundaries connected to the same magnetic field line will move apart
  from each other in a random walk fashion due to random photospheric motion.
Therefore, a typical field line will generally be stretched by the separation of
  the footpoints and the magnetic energy will increase.
This increase in the magnetic energy must come from the Poynting flux.
We see from Figure~\ref{ts}(d) that $\bar{I}$ tends to saturate in the long-time limit
at a level close to that of $\bar{W}$.
In principle, these two rates should be the same, since the time averaged total energy 
  also tends to a constant level.
Numerically there is a slight difference between the two. 
Convergence studies show that this is mainly due to inaccuracy from finite
grid size, and the difference decreases when higher resolution is used,
especially in the parallel direction.

Another measure of the accuracy of our solutions is to test the energy balance equation,
\begin{equation}
  \frac{ d (E_M + E_K)}{dt} = I - W_\eta - W_\nu
\label{eng_bal}\end{equation}
Figure~\ref{ts} (e) shows $I$ as a function of time in pink for the run R5, 
  $d(E_M + E_K)/dt$ (calculated by taking finite difference in time) 
  in purple, $-W_\eta - W_\nu$ in blue,
  and the difference between the right and left hand sides of Equation~(\ref{eng_bal})
  in green.
We do see that the residual power due to numerical inaccuracy is generally small
  compared with other terms.
While accuracy can be improved by running at higher resolutions,
  doing so would require much longer run times,
as well as limit the highest Lundquist number that can be simulated.  
In the context of energy balance in our simulations, we remark that the energy
dissipated due to ohmic or viscous terms is essentially converted into thermal
energy.
No energy term or transport equations are included.
This is perhaps the primary weakness of this model, 
  as it prevents us from  predicting temperature and density profiles 
  which can be directly compared with observations (see \citealt{Dahlburg2009}). 
However, the heating rate required to maintain observed coronal temperatures
  can indeed be estimated as has been done in, e.g., \cite{PHT2002}.
NB08 followed this practice and found that the heating rate
  determined from 2D simulations is consistent with such estimation, 
if the energy dissipation does turn into heat as assumed.
Readers should compare similarities and differences between this 
  treatment with those used in other studies
  (\citealt{Long1993PhD,ls1994,Rappazzo,Hendrix1996a,Hendrix1996b,GalsNord1996,
    Rappazzo2010a}). 
    
Figure~\ref{ts} (f) shows $\bar{B}_{\perp}$ as a function of time.
Here $\bar{B}_{\perp}$ is defined as a root-mean-square value of the magnetic field
strength, and so is effectively the square root of $E_M$ per unit volume.
Similar to other time-averaged quantities, $\bar{B}_{\perp}$ also tends to
saturate at long time, at values that are much more physically reasonable than
those in the 2D runs. 
We remark that these saturated levels in
the 3D runs are already much more reasonable than those in the 2D runs that
were found to have a scaling of 
   $\bar{B}_{\perp} \propto \eta^{-1/2}$,
   which can be much larger than unity (the value of the constant $B_z$ used 
   in the simulations).
Thus, inclusion of 3D effects is seen to reduce the magnitude of $\bar{B}_{\perp}$ to values
that are smaller than the magnitude of $B_z$.

These effects include the formation of thin current layers, 
onset of instabilities, and subsequent reconnection and enhanced energy dissipation.
All these effects are more prominent when $B_{\perp}$ is larger, 
effectively limiting the growth of $B_{\perp}$.
Thus, these 3D effects can self-regulate the level of $B_{\perp}$ that can be built up 
  when subjected to the driving of the random footpoint motion.
  
Due to the fact that we are injecting energy into the system through random 
  photospheric footpoint motion,
a natural question to ask is whether this would induce other random processes,
such as a turbulent cascade of energy that contributes to the heating of the corona.
Indeed, turbulence has been studied in various coronal loop heating models
beginning with the early work of \citet{vb1986} and more
recently by others (\citealt{Hendrix1996b,Dmitruk1998,Rappazzo}). 
However, as mentioned in the above discussion, we are driving with slow 
  boundary flows (less than 1/10 of the \al speed) with $\tau_c \gg \tau_A$,
  and thus the magnetic field configuration maintains quasi-equilibrium most 
  of the time.
Moreover, we apply random boundary flows,
instead of constant motion as in, e.g., \cite{Rappazzo}, 
 so that energy injection is much slower due to the fact that magnetic field lines are 
 stretched in a random walk fashion rather than at a constant rate.  
As a result, we have $E_K \ll E_M$, which is not consistent with
  equipartition of energy in \al wave turbulence.
Similar to \cite{Hendrix1996b}, energy
  spectra in our simulations are largely exponential (not shown here) 
  during relatively quiescent periods,
with little or no impulsive energy release, 
but become progressively shallow power laws during particularly
  intense current sheet disruption events,
  with possible excitation of more \al waves for a short duration.
As with their study however, computation grid resolutions
  enable us to resolve less than a decade of the inertial range
  of the energy cascade. 
  
While it seems turbulence plays just a minor role in our present analysis, 
whether it plays a crucial role in
  determining the speed of magnetic reconnection has attracted a number
  of recent investigations (e.g. 
  \citealt{Lazarian1999,Loureiro2007,Bhattacharjee2009,Loureiro2009,Kowal2009}).
It is evident there is a surge of interest in numerical experiments concerning
  turbulent reconnection and that much has yet to be settled. 
It would be interesting to see if any insights can be gleaned from our own data.  
As mentioned above, the presence of turbulence seems to be intermittent in our simulations, 
mainly during intense impulsive current sheet disruption events. 
Of crucial importance here is how well resolved we can be, 
and therefore how well developed an inertial range we can identify.
This will depend on how low the value of $\eta$ (and thus how high the Lundquist number $S$) 
  we can simulate in 3D,
as well as how important physical properties scale with $\eta$ or $S$,
which we turn our attention to now.

Figure~\ref{3dplots} shows some of the scaling results we have obtained so far. 
In Figure~\ref{3dplots} (a), the time-averaged ohmic dissipation rate $\bar{W}_\eta$ 
  (at the saturated level) for different $\eta$ for the runs listed
  in Table~\ref{tbl-1} is plotted in triangles,
while the viscous dissipation rate $\bar{W}_\nu$ are plotted in squares.
As pointed out above, $\bar{W}_\nu \ll \bar{W}_\eta$ in general, 
and thus the total dissipation rate (heating rate) 
  $\bar{W} = \bar{W}_\eta + \bar{W}_\nu$ (plotted in asterisks)
  is very close to $\bar{W}_\eta$, except in the large resistivity limit.
The time-averaged Poynting flux $\bar{I}$ is also plotted in the same graph in circles.
It should be the same value as $\bar{W}$ theoretically, 
and we do see that the differences between these two quantities 
  are generally small in our numerical results,
  indicating acceptable accuracy.
  
From this plot, we see that $\bar{W}$ actually only changes within an order of magnitude,
and levels off at both the large and small $\eta$ limit.
This has important implications for the coronal heating problem,
  since the Lundquist number
  (on the order of the inverse of the normalized $\eta$ in our simulations)
can be as high as $10^{14}$ in the solar corona.
Therefore, the leveling off of  $\bar{W}$ at the small $\eta$ limit is especially important,
  and is in fact predicted by NB08 based on 2D simulations and 
  theoretical arguments.
As mentioned above, this level of $\bar{W}$ was shown in NB08 to be
  independent of dissipation mechanism provided that the correlation time $\tau_c$
  is small compared with the time to build up magnetic energy.
It was also estimated that this level of heating rate can give 
  the same order of magnitude required for realistic coronal heating,
following similar considerations as in \cite{PHT2002}.
However, the amount of magnetic energy built up in this process does
  depend on dissipation mechanism,
and becomes unphysically large in 2D simulations in the small $\eta$ limit
  (with $\bar{B}_{\perp}$ scaling as $\eta^{-1/2}$).
We will now show that this scaling becomes much weaker in 3D.

Figure~\ref{3dplots} (b) shows the time-averaged $\bar{B}_{\perp}$ 
  (at the saturated level) for different $\eta$.
This is a measure of the magnetic field (or magnetic energy) production
  in the statistical steady state due to the applied random photospheric motion.
Unlike $\bar{W}$, $\bar{B}_{\perp}$ production changes over an order of 
  magnitude from large to small $\eta$.
This is because in the highly resistive limit, 
  magnetic field produced is quickly dissipated and can only reach a low magnitude,
  while the dissipation rate does not decrease that much.
In the small resistivity limit, the increase of $\bar{B}_{\perp}$ slows down significantly.

Due to the fact that we are doing 3D simulations,
and that we need to simulate for a long time to obtain good statistics,
so far we have only been able to extend the value of $\eta$ to about an
  order of magnitude lower, as compared with similar studies in  \citep{ls1994}.
Nevertheless, we can see already that below $\eta \sim 10^{-3}$, 
  there is a significant deviation 
  from the scalings obtained in \cite{ls1994}, who showed by numerical computation and scaling analysis 
  that both $\bar{W}$ and $\bar{B}_{\perp}$ should scale with $\eta^{-1/3}$
  in the small $\eta$ limit.
We have added dotted lines in Figure~\ref{3dplots} (a) and (b) showing 
  the $\eta^{-1/3}$ scaling.
We see that the portion of the data in a range close to $\eta \sim 10^{-3}$
  is indeed consistent with an $\eta^{-1/3}$ scaling.
However, as described above, both $\bar{W}$ and $\bar{B}_{\perp}$ increase 
  much slower with decreasing $\eta$ for even smaller $\eta$.
This result has important implications on the solar coronal heating problem,
since the Lundquist number in the solar corona is very high and hence 
  we most likely need to have a mechanism to provide coronal heating that is 
  independent of the Lundquist number in order to obtain a physically reasonable heating rate. 
At the same time, the magnetic field energy production should not increase to
  an unreasonable level compared with observations.
In addition to this numerical evidence, we will provide our own scaling analysis 
  to make sense of these results, 
as well as compare with results in \cite{ls1994}.

\section{Scaling Analysis} \label{LongFlx}

We have shown an initial confirmation of the hypothesis of NB08 but as an
  additional goal we would like to understand the exact mechanism 
  giving rise to saturation. 
Their conjecture made clear that the insensitivity
  to $\eta$ holds true no matter what the saturation mechanism is. 
In order to provide a more complete numerical confirmation of their conjecture, 
it becomes necessary to identify the possible physical mechanisms behind saturation. 

A natural place to begin would be to examine the
  results of \cite{Long1993PhD} and \cite{ls1994} who derived scaling laws based on
  Sweet--Parker reconnection theory and analyzed a range in $\eta$ which we have
  covered in our own study. 
By looking at  where their scaling behavior or where
  their assumptions might be failing in our own numerical results, 
we might gain some insight into the physics
  occurring at even lower $\eta$. 
The reader is referred to these papers for a detailed review of their scaling
  arguments. 
Here we will only discuss their assumptions and results briefly.  

They assumed Sweet--Parker theory is valid in the sense that
  when looking at only the
  current sheet region forming between two coalescing islands (flux tubes), 
the reconnection can be treated as a steady process in
  resistive MHD which results in the classic Sweet--Parker scaling
  relating the width $\delta$ and length $\Delta$ of a reconnecting
  current sheet: 
\begin{equation} 
   \delta / \Delta \sim S_{\perp}^{-1/2} 
\label{sp-scaling}\end{equation}
where  $S_{\perp}\equiv\bar{B}_{\perp} w / \eta$ is the perpendicular Lundquist number,
with $w$ being the perpendicular length scale of the reconnecting islands and so
  $w \sim v_p  \tau_c $ with $v_p$ being the root-mean-square value of the random 
  photospheric flow velocity.
They also observed that both the number of current sheets $N$ in the simulation box and the length
  of the current sheets $\Delta$ are relatively insensitive to resistivity. 
We follow these assumptions as a starting point for our discussion,
  although we recognize that some of them need to be re-examined more carefully
  in future studies.
  
In particular, the Sweet--Parker reconnection theory should apply only to 
  higher Lundquist number (smaller $\eta$) cases, 
in which the energy dissipation is dominated by the reconnection process.
Therefore, the scaling analysis presented here should not work 
  for larger $\eta$ (i.e. $\eta > 0.01$ here), which is actually not within 
  the focus of our studies here.
When the energy dissipation is mainly from the Sweet--Parker 
  current sheets, the dissipation rate can be estimated by
\begin{equation} 
   \bar{W} \sim \eta N \Delta L \frac{\bar{B}^2_\perp}{\delta} \sim 
   \frac{\bar{B}^2_\perp L L^2_\perp}{\tau_E}
\label{wbar-cs}\end{equation}
where we have used the estimation that the current density of the current sheet
  is given by $J \sim \bar{B}_{\perp}/\delta$ and that the volume of the simulation
  box is $L L^2_\perp$.
The energy dissipation timescale $\tau_E$ in Equation~(\ref{wbar-cs}) can then be solved as
\begin{equation} 
   \tau_E \sim L^2_\perp / N (\eta w \bar{B}_\perp)^{1/2}
\label{wbar-tauE}\end{equation}
where we have used the Sweet--Parker scaling in Equation~(\ref{sp-scaling}).
In a statistical steady state, the energy dissipation by Equation~(\ref{wbar-cs}) has to be
  replenished by the production of magnetic field energy 
  due to the footpoint motion within the same amount of time $\tau_E$.

In the studies of \cite{Long1993PhD} and \cite{ls1994}, 
although random photospheric motion was used in the simulations,
the effects due to such random flows were not taken into account in their
  scaling analysis.
This can be justified if $\tau_c$ is much larger than the energy dissipation time $\tau_E$.
In this case, the magnetic field strength production is given by
\begin{equation} 
   \bar{B}_\perp \sim B_z \frac{v_p \tau_E}{L} 
   \sim \left[ \left( \frac{B_z v_p}{L N}\right)^2 \frac{L^4_\perp}{w \eta} \right]^{1/3} 
\label{Bbar-long}\end{equation}
where we have used Equation~(\ref{wbar-tauE}) and solved for $\bar{B}_\perp$.
Putting back Equation~(\ref{Bbar-long}) into Equation~(\ref{wbar-tauE}) results in
\begin{equation} 
   \tau_E \sim \left( \frac{L^4_\perp L}{N^2 w B_z v_p \eta}\right)^{1/3} 
\label{tauE-long}\end{equation}
and so the energy dissipation rate becomes
\begin{equation} 
   \bar{W} \sim  \left( \frac{L^{10}_\perp B_z^5 v_p^5}{L^2 N^2 w \eta}\right)^{1/3} 
\label{wbar-long}\end{equation}
after putting Eqs.~(\ref{Bbar-long}) and (\ref{tauE-long}) into Equation~(\ref{wbar-cs}).
Note that all three of these quantities, $\bar{B}_\perp$, $\tau_E$, 
and $\bar{W}$ scale with $\eta^{-1/3}$,
and thus we have recovered scaling laws derived in \cite{Long1993PhD} and \cite{ls1994}, 
although we are using a slightly different approach.

We may now put reasonable numbers into 
  Eqs.~(\ref{Bbar-long}) to (\ref{wbar-long}) 
  and compare with our simulation results.
Our simulations are set up to use $L = L_\perp = B_z = 1$.
The root-mean-square photospheric flow velocity is measured numerically to be 
  $v_p \sim 0.075$, and thus $w \sim v_p \tau_c = 0.75$.
The average number of current sheets $N$ is more difficult to determine 
  and is subject to some uncertainties. 
However, we have done some analysis (not shown here) 
  of our simulations for different $\eta$
  and found that $N \sim 7$ numerically in the small $\eta$ limit.
This seems to be somewhat higher than expected from the number of
  reconnecting islands (flux tubes).
However, it is actually quite common to see multiple current sheets 
  in a simulation output, as shown in Figure~\ref{cs}.
  
\begin{figure}
\plotone{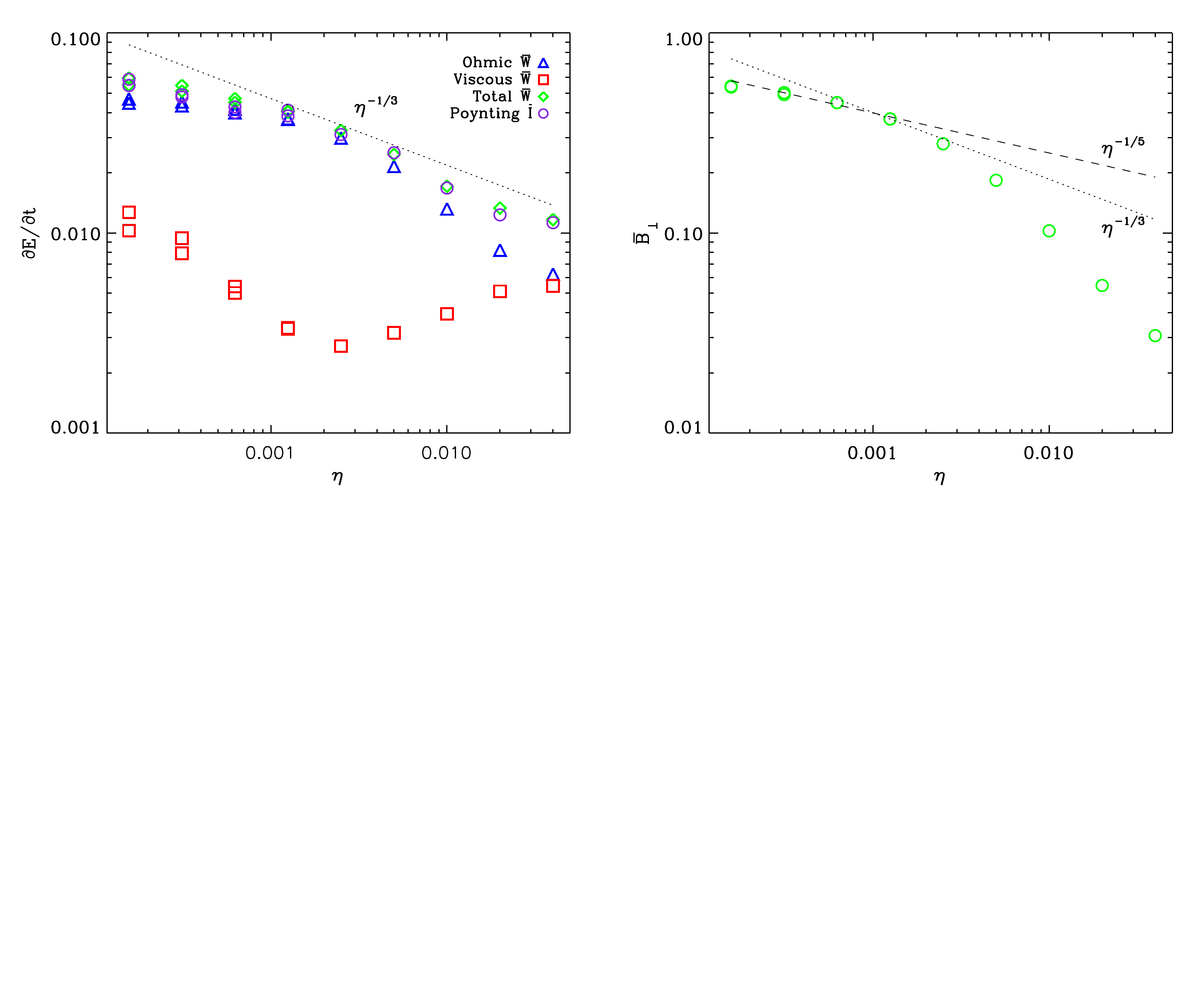}
\caption{(a) Average energy dissipation rate for different values of $\eta$. $\triangle$ is ohmic dissipation, $\square$ is viscous dissipation, $\Diamond$ is the total of the two, and $\bigcirc$ is the footpoint Poynting flux. 
(b) Average perpendicular magnetic field strength for different values of $\eta$.}
\label{3dplots}
\end{figure}

Based on these values, we have $\tau_E \sim 0.71/\eta^{1/3}$, 
  and so $\tau_E \sim 7.1$ for $\eta = 10^{-3}$.
At the same time we get $\bar{B}_\perp \sim 0.53$ from Equation~(\ref{Bbar-long}),
and $\bar{W} \sim 0.04$ from Equation~(\ref{wbar-long}) at the same $\eta$.
Both of these are close enough to the values found in 
  Figure~\ref{3dplots} (a) and (b), and so it is an indication that our parameters
  used in these estimations are consistent with simulations.
Compared with the value of $\tau_c = 10$, we see that although $\tau_E$
  is still smaller than $\tau_c$, it is about the same order of magnitude and
  thus Equation~(\ref{Bbar-long}) is only marginally justified.
For larger $\eta$, $\tau_E$ is smaller, e.g., $\tau_E \sim 3.3$ for $\eta = 0.01$ and thus
  is much smaller than $\tau_c$ so that the random effect is not as important.
This qualitatively explains why we see from Figure~\ref{3dplots} (a) and (b)
  that there is a range roughly around $\eta \sim 0.01$ to 0.001 where 
  both $\bar{W}$ and $\bar{B}_\perp$ scale approximately as $\eta^{-1/3}$,
  as indicated by the two dotted lines in the two plots.
However, for smaller $\eta$, $\tau_E$ becomes larger,
e.g., $\tau_E \sim 15$ for $\eta = 10^{-4}$ 
(if we continue to use the approximation, which may not be strictly valid),
and thus larger than $\tau_E$ which makes the effect of randomness important.
This explains the deviation from the $\eta^{-1/3}$ scaling for both 
  $\bar{W}$ and $\bar{B}_\perp$ for $\eta$ smaller than around $10^{-3}$.
  
Now, taking into account the effect of random boundary flow, 
  which makes the footpoints move in a random walk fashion 
  as argued in NB08,
  the estimate for magnetic field production must be changed 
  from Equation~(\ref{Bbar-long}) to
\begin{equation} 
   \bar{B}_\perp \sim B_z \frac{v_p (\tau_c \tau_E)^{1/2}}{L} 
   \sim \left[ \left( \frac{B_z v_p L_\perp}{L}\right)^4 \frac{\tau_c^2}{N^2 w \eta} \right]^{1/5} 
\label{Bbar-ran}\end{equation}
where we have again used Equation~(\ref{wbar-tauE}) and solved for $\bar{B}_\perp$.
Substituting Equation~(\ref{Bbar-ran}) back into Equation~(\ref{wbar-tauE}) results in
\begin{equation} 
   \tau_E \sim \left[ \frac{L^8_\perp}{N^4 \tau_c} \left(\frac{L}{w B_z v_p \eta} \right)^2 \right]^{1/5} 
\label{tauE-ran}\end{equation}
and so the energy dissipation rate becomes
\begin{equation} 
   \bar{W} \sim  \frac{L^2_\perp}{L} B_z^2 v_p^2 \tau_c 
\label{wbar-ran}\end{equation}
after putting Eqs.~(\ref{Bbar-ran}) and (\ref{tauE-ran}) into Equation~(\ref{wbar-cs}), and thus 
  it is independent of $\eta$.
Note that Equation~(\ref{wbar-ran}) is exactly the same as found in NB08
  for systems regardless of dissipation mechanism,
and is estimated to give the same order of heating consistent with observations.

Using the same values of $L$, $L_\perp$, $B_z$, $\tau_c$, $v_p$, $w$, 
and $N$, Equation~(\ref{tauE-ran}) becomes $\tau_E \sim 0.42/\eta^{2/5}$, 
  and thus $\tau_E \sim 6.7$ for $\eta = 10^{-3}$, 
if we could apply this equation. 
This turns out to be very close to $\tau_E \sim 7.1$ 
  estimated above using Equation~(\ref{tauE-long}),
  indicating that the transition point between these two regimes of scalings
  is around $\eta = 10^{-3}$ in our simulations.
For $\eta = 10^{-4}$, Equation~(\ref{tauE-ran}) gives $\tau_E \sim 17$,
  which is significantly larger than $\tau_c$, and so these scalings based
  on random walk of footpoints are justified.

Based on this set of parameters, Equation~(\ref{wbar-ran}) predicts $\bar{W} \sim 0.056$
  (independent of $\eta$), which is close to the asymptotic values found in 
   Figure~\ref{3dplots} (a) in the small $\eta$ limit.
We do see from this plot that $\bar{W}$ indeed does not increase as fast 
  when $\eta$ is below $10^{-3}$,
and is consistent with a trend to a constant level in small $\eta$,
although we still only have a limited range of $\eta$ that we can simulate.
At the same time, Equation~(\ref{Bbar-ran}) gives a value of $\bar{B}_\perp \sim 0.97$,
which is somewhat larger than expected from Figure~\ref{3dplots} (b),
although we do need to recognize that there are uncertainties in these
  scaling estimates.

A better test of Equation~(\ref{Bbar-ran}) would be the scaling with $\eta$ in the
  small $\eta$ limit.
In Figure~\ref{3dplots} (b), we have also plotted a dashed line indicating the 
  scaling of $\eta^{-1/5}$.
We do see that this seems to be consistent with a portion of the data 
  of $\bar{B}_\perp$ below $\eta \sim 10^{-3}$. 
However, we cannot rule out the possibility that $\bar{B}_\perp$ is 
   actually increasing slower than $\eta^{-1/5}$, 
   possibly due to a modification of the Sweet--Parker reconnection
   scalings, e.g., Equation~(\ref{sp-scaling}).
We will further discuss this possibility in the next section.
    
\section{Discussion and Conclusion} 
\label{fwcr}

In this paper, we have presented our latest results based on numerical simulations
  of a 3D RMHD solar corona heating model for a range of $\eta$ (and thus
  Lundquist number) with random photospheric motion.
These simulations were performed over a period of more than two years
  and numerical results have been verified carefully to eliminate possible errors.
So far, we have been able to simulate cases with $\eta$ about an order of
  magnitude smaller than those presented in similar studies in 
  \citep{Long1993PhD} and \citep{ls1994}.
While this extension seems modest, it actually requires much more computational
  effort due to the increase in resolution and running time required, 
  as well as the decrease of time-step for numerical stability.
To be able to achieve that, we have been running our simulations in parallel computers,
  as well as using GPUs.
  
Moreover, we have shown that the extension of this scaling study towards
  smaller $\eta$ turns out to have very important physical consequences.
Numerically, we have shown that the scaling laws 
  (with $\bar{W}$ and $\bar{B}_\perp$ scale with $\eta^{-1/3}$) found in  
  \cite{Long1993PhD} and \cite{ls1994} become invalid for $\eta$ 
  smaller than what was used in their studies (around $\eta \sim 10^{-3}$).
Both $\bar{W}$ and $\bar{B}_\perp$ are now found to increase much more slowly
  for smaller $\eta$, with $\bar{W}$ possibly leveling off to an asymptotic value.
We have presented our own scaling analysis to justify our numerical results.
  By following similar assumptions as in \cite{Long1993PhD} and \cite{ls1994},
  e.g., using Sweet--Parker scalings, we have been able to recover their
$\eta^{-1/3}$ scaling laws for a range of $\eta$ larger than $10^{-3}$.
We have demonstrated that the transition between scaling behaviors derives from the fact
  that the effects of random photospheric motion are not important in the larger
  $\eta$ range where the energy dissipation time $\tau_E$ is smaller than 
  the correlation time $\tau_c$ of the random flow.
For $\eta$ smaller than around $10^{-3}$,
  $\tau_E$ becomes comparable or even larger than $\tau_c$.
In this range, an analysis based on the random walk of photospheric
footpoints motion will predict the insensitivity to $\eta$ we observe,
further substantiating the results found in NB08 which were based
on 2D simulations and more general theoretical considerations. 
This is important to the problem of coronal heating since this heating
  rate has been shown to be consistent with the requirements for 
  coronal heating.
  
We have also shown that now $\bar{B}_\perp$ has a much weaker scaling
   with $\eta$, i.e., $\eta^{-1/5}$ instead.
This is much better than the $\eta^{-1/2}$ scaling in 2D simulations, 
  as well as weaker than the $\eta^{-1/3}$ scaling found in 
  \cite{Long1993PhD} and \cite{ls1994}.
This means that this scaling will predict a more physically realistic level
  of magnetic field as compared with observations.
However, due to the fact that the Lundquist number of ($\sim$ inverse of $\eta$) 
  the solar corona can be very high (up to $10^{12}$ -- $10^{14}$), 
even a $\eta^{-1/5}$ scaling would result in an unrealistically large magnetic field,
  despite a much weaker dependence.
The reason behind this is the fact that the Sweet--Parker reconnection rate,
  which scales with $\eta^{1/2}$ is too slow for high Lundquist numbers.
  
One solution for this problem is the possibility of a higher rate of 
  magnetic reconnection even under resistive MHD.
This possibility has attracted a number of recent investigations (e.g., 
  \citealt{Lazarian1999,Loureiro2007,Bhattacharjee2009,Loureiro2009,Kowal2009}).
Many of these studies fall within the scope of turbulent reconnection.
While there are some indications that $\bar{B}_\perp$ found in our simulations
  might actually scale weaker than even $\eta^{-1/5}$,
  we still have not been able to simulate even smaller $\eta$ to 
  confirm this definitively.
Moreover, the effects due to turbulence are still too difficult to study
  using our current level of resolution.
However, this question is important enough that we are trying different
  ways to extend our range of $\eta$ to even smaller values 
  to study these effects.
  
In summary, by simulating with $\eta$ about an order of magnitude smaller
  than in previous studies,
we have been able to find new physical effects due to random photospheric
  flows and thus new scalings with Lundquist number.
In future work, with further improvements in our computational approach, we
hope to report results with even smaller values of $\eta$, and investigate the
possibilities of another asymptotic range involving secondary instabilities
and turbulent processes.

  


\acknowledgments

The authors thank Matt Gilson for his contribution
in programming a part of the code as a summer student project.
Computer time was provided by UNH (using the Zaphod Beowulf 
cluster at the Institute for the Study of Earth, Oceans and Space), as well as grants of HPC
resources from the Arctic Region Supercomputing Center, the DoD High
Performance Computing Modernization Program, and NSF TeraGrid resources
provided by NCSA. This research is supported in part by grants from  NSF (AGS-0962477 and
AGS-0962698), NASA (NNX08BA71G, NNX09AJ86G, and NNX10AC04G) and DOE (DE-FG0207ER46372).

\clearpage



\tabletypesize{\scriptsize}
\begin{deluxetable}{lllllllcll}
\tablewidth{0pt}
\tablecaption{Summary of Numerical runs\label{tbl-1}}
\tablehead{
\colhead{Run} & \colhead{$\eta$} & \colhead{$\nu$} & \colhead{$B_{\perp}$} & \colhead{$S_{\perp}$} &
\colhead{$W_{\eta}$} & \colhead{$W_{\nu}$} & \colhead{Poynting} &
\colhead{$T/\tau_{A}$} & \colhead{Resolution}
}
\startdata
R0 & 0.00015625 & 0.00015625 & 0.542 & 3470 & 0.0446 & 0.0103 & 0.0587 &  305.545 & 1024$^{2}\times$128 \\
R1 & 0.00015625 & 0.00015625 & 0.537 & 3440 & 0.0468 & 0.0127 & 0.0546 &  487.252 & 512$^{2}\times$64 \\
R2 & 0.00015625 & 0.00062500 & 0.610 & 3900 & 0.0513 & 0.0283 & 0.0519 &  245.546 & 512$^{2}\times$32 \\
R3 & 0.00015625 & 0.00062500 & 0.614 & 3930 & 0.0498 & 0.0275 & 0.0458 &  77.0269 & 512$^{2}\times$32 \\
R4 & 0.00031250 & 0.00031250 & 0.492 & 1570 & 0.0433 & 0.00792 & 0.0491 &  857.407 & 512$^{2}\times$64 \\
R5 & 0.00031250 & 0.00031250 & 0.503 & 1610 & 0.0452 & 0.00941 & 0.0478 &  9321.75 & 256$^{2}\times$32 \\
R6 & 0.00031250 & 0.00062500 & 0.502 & 1610 & 0.0431 & 0.0111 & 0.0467 &  2032.02 & 256$^{2}\times$32 \\
R7 & 0.00062500 & 0.00062500 & 0.449 & 718 & 0.0416 & 0.00540 & 0.0427 &  19342.8 & 128$^{2}\times$32 \\
R8 & 0.00062500 & 0.00062500 & 0.448 & 717 & 0.0399 & 0.00502 & 0.0401 &  820.339 & 128$^{2}\times$32 \\
R9 & 0.0012500 & 0.0012500 & 0.372 & 298 & 0.0370 & 0.00332 & 0.0385 &  11668.2 & 128$^{2}\times$32 \\
R10 & 0.0012500 & 0.0012500 & 0.371 & 297 & 0.0373 & 0.00336 & 0.0411 &  706.141 & 64$^{2}\times$16 \\
R11 & 0.0025000 & 0.0025000 & 0.279 & 112 & 0.0299 & 0.00272 & 0.0311 &  1317.70 & 64$^{2}\times$16 \\
R12 & 0.0050000 & 0.0050000 & 0.183 & 36.7 & 0.0215 & 0.00317 & 0.0252 &  2566.96 & 64$^{2}\times$16 \\
R13 & 0.0100000 & 0.0100000 & 0.103 & 10.3 & 0.0132 & 0.00394 & 0.0168 &  5209.60 & 64$^{2}\times$16 \\
R14 & 0.020000 & 0.020000 & 0.0547 & 2.73 & 0.00822 & 0.00511 & 0.0123 &  10245.4 & 64$^{2}\times$16 \\
R15 & 0.040000 & 0.040000 & 0.0307 & 0.767 & 0.00623 & 0.00544 & 0.0113 &  10240.3 & 32$^{2}\times$64 \\
R16 & 0.080000 & 0.080000 & 0.0197 & 0.246 & 0.00550 & 0.00612 & 0.0105 &  10240.5 & 32$^{2}\times$64 \\
\enddata
\end{deluxetable}

\clearpage

\clearpage

\clearpage

\end{document}